\documentclass[twocolumn,showpacs,amsmath,amssymb,floatfix,superscriptaddress]{revtex4}
\usepackage{graphicx}
\usepackage{dcolumn}
\usepackage{bm}
\usepackage{hyperref} 
\usepackage{latexsym}

\begin{document}
\title{Does a single zealot  affect an infinite group of voters ?}
 \author{Mauro Mobilia}
 \affiliation{
Laboratoire d'Opto\'electronique des  Mat\'eriaux Mol\'eculaires (LOMM), Institut des Mat\'eriaux (IMX), 
Ecole Polytechnique F\'ed\'erale de Lausanne, CH-1015 Lausanne  EPFL,
 Switzerland.
\\
and
\\
 Center for BioDynamics,
  Center for Polymer Studies \&  Department of Physics, Boston University,
  Boston, MA, 02215 
}

\begin{abstract}
A method for studying exact properties of  a class of {\it inhomogeneous}
stochastic many-body systems is developed and presented in the framework of
 a voter model perturbed by the presence of a ``zealot'', an individual
 allowed to favour an opinion.
We compute exactly the magnetization of this model and find that 
in one ($1d$) and two dimensions ($2d$)
 it evolves, algebraically  ($\sim t^{-1/2}$) in $1d$ and much 
slower ($\sim 1/\ln{t}$) in $2d$, towards
 the unanimity state chosen by the zealot. In
 higher dimensions the stationary magnetization is no longer 
uniform: the zealot cannot influence all the individuals. Implications to
 other physical problems are also pointed out.
\end{abstract}
\maketitle 
\date{\today}

Researchers have devoted much attention to the field of non-equilibrium
 many-body stochastic processes \cite{Privman}. In particular the study
 of exact solutions of prototypical models such as
 the {\it voter model} \cite{Liggett} has proven to be fruitful
 for understanding a
 large class of  non-equilibrium phenomena \cite{Privman}.
However, more realistic systems taking into account the important dynamical 
effects of inhomogeneities, constraints and
 disorder (see e.g. \cite{Privman,dis}  and references  therein)
are less understood. 
To gain comprehension  of these situations, exact results for systems
 modeling inhomogeneous situations in any dimension are desirable,
 but rather scarce \cite{dis}.

In this work, motivated by these considerations, and with the 
voter model as a paradigm, we present techniques for computing 
 exact properties (in any dimension) of a class of 
stochastic many-body systems with {\it inhomogeneities}.

The voter model is an Ising-like model where an ``individual'' (or spin) 
associated to a lattice site ${\bf r}$
can have two different opinions $\sigma_{\bf r}=\pm 1$ \cite{Liggett}. The
dynamics of the system is implemented by randomly chosing one individual and assigning to it the
value of the spin of one of its randomly chosen nearest neigbors. In the
voter model, the global magnetization 
is conserved and the dynamics is $Z_2$ symmetric (invariance under the global
inversion $\sigma_{\bf r}\rightarrow -\sigma_{\bf r}$). 
 The importance of the voter model stems from the fact that it is one of a very few stochastic many-body systems that are solvable in any dimension and is
 useful for describing the kinetics of catalytic reactions
 \cite{K1,Frachebourg,Ben},
 in studying coarsening phenomena \cite{Ben,Dornic} and as a prototype
 model of opinion dynamics \cite{Vazquez}.

For the sake of concretness, and without loss of generality, we specifically
present our method and techniques in the framework of
  an {\it inhomogeneous voter model} where, 
to mimic in a simple manner the fact that a group of agents may
 have heterogeneous interactions,
 the conventional voter dynamics is supplemented by 
the presence of a ``zealot'': a
biased individual who favours one opinion. 
 We study the effect of this perturbation by computing the exact 
long-time magnetization,
 which is no longer conserved, in dimensions $d=1, 2$ and $3$. 
In low-dimensions the zealot drastically affects the dynamics: the system 
evolves towards  unanimity with the latter. The approach to the stationary 
state is algebraic in $1d$ and logarithmically slow in $2d$. In $d\geq 3$,
 the effect of the zealot is less dramatic and the local
 stationary magnetization is a  non-trivial function of
the distance to this biased individual. These findings, although formulated
in an ``opinion dynamics'' language, are, as pointed out hereafter,
 relevant to other physical problems.

The inhomogeneous voter model that we study is defined on 
a hypercubic lattice of size $(2L+1)^d$, where
 individuals, labelled by a vector ${\bf r}$ having components $-L \leq r_i\leq L$ (with $ i=1,\dots,
 d$),  may interact according to the usual voter dynamics.
 In addition, we now consider that a zealot, at 
site ``{\bf 0}'', tends to favour the diffusion 
of the  $+1$ opinion via the interaction with his
 neighbors: the zealot is the only individual in
 the system allowed to change his state from $-1$ to $+1$ (with rate $\gamma>0$) without regards to his neighbors, with whom he nevertheless interacts.   
According to the spin formulation of the model, the state of the system 
is described by the collection of all
 spins: $S\equiv \{\sigma_{\bf r}\}$. In this language, the
 dynamics of the  model is  governed 
by the usual voter model transition-rate \cite{Privman,K1,Frachebourg,Liggett} supplemented by a local 
term involving the local zealot's reaction. The spin-flip rate, $w_{\bf r}(S)\equiv 
w(\sigma_{\bf r}\rightarrow -\sigma_{\bf r} )$, therefore reads 
\begin{eqnarray}
\label{SF}
w_{\bf r}(S)=
\frac{1}{\tau}\left(1-\frac{1}{2d}\sigma_{\bf r}\sum_{{\bf r'} } 
\sigma_{\bf r'} \right)
+ \frac{\gamma}{2}\left
( 1- \sigma_{\bf 0}\right)\delta_{{\bf r},{\bf 0}}.
\end{eqnarray} 
Here the sum on right-hand side (r.h.s.) runs over the $2d$ nearest neighbors ${\bf r'}$ 
of site ${\bf r}$ and $\tau\equiv 1/\beta d >0$ defines the time scale.
The probability distribution $P(S,t)$ satisfies the master equation:
\begin{eqnarray}
\label{master}
\frac{d}{dt} P(S,t)= \sum_{\bf r}\left[
 w_{\bf r}(S^{\bf r}) P(S^{\bf r} ,t) - 
w_{\bf r}(S) P(S,t) \right],
\end{eqnarray}
where the state  $S^{\bf r}$ differs from  
 $S$ only by the spin-flip of $\sigma_{\bf r}$.
With the master equation (\ref{master}), in the limit $L\rightarrow \infty$,  
the equation of motion of the local magnetization $S_{\bf r}(t) \equiv \sum_{S} 
\sigma_{\bf r} \,P(S,t)$ reads: 
\begin{eqnarray}
\label{MS}
\frac{d S_{\bf r}(t) }{d(\beta t)}=
\Delta_{\bf r}S_{\bf r}(t)
- \frac{\gamma}{\beta} \left(S_{\bf 0}(t) -1 \right)\delta_{{\bf r}, {\bf 0}}.
\end{eqnarray}

Here $\Delta_{\bf r}$ denotes the discrete Laplace operator:
$\Delta_{\bf r}S_{\bf r}(t)\equiv
 - 2d S_{\bf r}(t) +\sum_{{\bf r'} } S_{\bf r'}(t)$.

The last term on the r.h.s. of Eq.(\ref{MS}) is due to the effect 
of the zealot and only appears in the case where ${\bf r}={\bf 0}$.

An important consequence of  Eq.(\ref{MS}) is that the only
 possible {\it uniform} final magnetization is $S_{\bf r}(\infty)=1$, i.e. the 
state favoured by the zealot.
However, in $d\geq 3$, as shown hereafter, the stationary 
magnetization profile
 turns out to be {\it non-uniform} (but isotropic). 

Using the properties of the modified Bessel functions of the  first kind, $I_{r}(t)$
 \cite{Abramowitz}, we obtain the  formal solution of Eq.(\ref{MS}):
\begin{eqnarray}
\label{eq.7}
 && S_{\bf r}(t)= \sum_{\bf k}  S_{\bf k}(0) 
\prod_{i=1}^{d} \left[e^{-2\beta t}I_{k_{i}-r_i}(2\beta t)\right]\nonumber\\
&+&\gamma  \int_{0}^{t}dt'  \left[1- S_{\bf 0}(t-t')\right]\prod_{i=1}^{d}
\left[e^{-2\beta t'}I_{r_{i}}(2\beta t')\right].
\end{eqnarray}
To obtain an explicit expression for the magnetization we solve 
the self-consistent integral equation (\ref{eq.7}) for $\bf{r}=\bf{0}$ and then 
 plug the result back  into (\ref{eq.7}). For this purpose it is
 useful to denote the Laplace transform of a product of Bessel functions
(multiplied by an exponential term)
\begin{eqnarray}
\label{eq.9}
{\hat I}_{{\bf r}}(s,\beta)\equiv \int_{0}^{\infty} dt\; e^{-st} \left[
e^{-2d\beta t}I_{r_1}(2\beta t)\dots I_{r_d}(2\beta t)\right].
\end{eqnarray}
This quantity can be rewritten in terms of Watson integrals, or ``lattice Green-functions'':
\begin{eqnarray}
\label{Wat}
{\hat I}_{{\bf r}}(s,\beta)\equiv \int_{-\pi}^{\pi}
 \frac{d^{d}{\bf q}}{(2\pi)^d}
\frac{e^{-i{\bf q.r}}}{s+2\beta[d-\sum_{i=1}^{d}\cos{q_i}]},
\end{eqnarray}
where ${\bf q}=(q_1,\dots,q_d)$ is a $d-$dimensional vector.

Laplace-transforming Equation (\ref{eq.7}), and using the convolution theorem, we obtain
 the following expression for the Laplace transform of
 the local magnetization:

\begin{eqnarray}
\label{eq.11} 
\hat{S}_{\bf r}(s) \equiv \int_{0}^{\infty} dt \, e^{-st} 
S_{\bf r}(t) =
\frac{\gamma {\hat I}_{\bf r}(s, \beta)}{s\left[1+\gamma{\hat I}_{\bf 0}
(s, \beta)\right]},
\end{eqnarray}
where, for technical simplicity, we consider the  zero initial magnetization state:
 $S_{\bf k}(0)=0$.

The exact expression for the long-time  magnetization is obtained by Laplace-inverting the $s\rightarrow 0$ expansion of Eq.(\ref{eq.11}) and paying due attention to the situations where the integrals (\ref{Wat})
 are divergent. In the sequel the norm of the vector ${\bf r}$ is denoted $r$.

We first consider the one-dimensional problem.
In this case, the quantity (\ref{eq.9}) reads \cite{Abramowitz}:
\begin{eqnarray}
\label{eq.13}  
{\hat I}_{\bf r}(s, \beta)\equiv {\hat I}_{ r}(s, \beta)&=& 
\frac{\left\{[\sqrt{s+4\beta}-\sqrt{s}] /
(2\sqrt{\beta})\right\}^{2r}}{\sqrt{s(s+4\beta)}}.
\end{eqnarray}

It should  be noticed that the $s\rightarrow 0$ behavior of (\ref{eq.13}) diverges as $s^{-1/2}$.

Laplace-inverting the  $s\rightarrow 0$  expansion of (\ref{eq.11}), together with the expression
 of (\ref{eq.13}) for $r=0$, we get ($\beta t\rightarrow \infty$):
\begin{eqnarray}
\label{eq.15.0} 
S_{0}(t) = 1-\frac{2}{\gamma}\sqrt{\frac{\beta}{\pi t}}.
\end{eqnarray}

In the more general case, where $r>0$, the local magnetization is obtained 
similarly  from (\ref{eq.11}) and (\ref{eq.13}) [in the limit $s \rightarrow
0$, with $r\sqrt{s}$ kept fixed]: 
\begin{eqnarray}
\label{eq.16} 
 S_{r}(t) \simeq 
{\rm erfc}\left( \frac{r}{2\sqrt{\beta t}}\right),
\end{eqnarray}
where ${\rm erfc}(z)$ is the usual complementary error function
 \cite{Abramowitz}.
The expression (\ref{eq.16}) is valid for $\beta t \gg 1$, and is 
particularly useful
 in the scaling limit where both $r\rightarrow \infty$ and
 $t\rightarrow \infty$, but where the ratio $r/\sqrt{\beta t}$ is finite.

When  $r$ is finite (i.e. $0<r<\infty $ ) and $\beta t \rightarrow \infty$, 
we obtain the following long-time behavior:
\begin{eqnarray}
\label{eq.18} 
 S_{r}(t)=1-\frac{r+(2\beta/\gamma)}{\sqrt{\pi \beta t}}
\end{eqnarray}
In two dimensions, the integral (\ref{Wat}) is also divergent 
in the long-time regime $s\rightarrow 0$ and therefore its main contribution arises from 
$ q^2\equiv q_1^2 +q_2^2 \rightarrow 0$.
In this sense, we first expand (\ref{Wat}) for small $s$
 in the case where ${\bf r}={\bf 0}$:
\begin{eqnarray}
\label{eq.20}  
{\hat I}_{ \bf 0}(s, \beta) \xrightarrow[s\rightarrow 0]{}
-\frac{1}{4\pi\beta}\ln\left[\frac{s}{\beta}\right], 
\end{eqnarray}
which implies that the long-time behavior of the average-opinion of the zealot is
\begin{eqnarray}
\label{eq.22} 
S_{\bf 0}(t) -  S_{\bf 0}(\infty)\simeq -
\left(\frac{4\pi\beta }{\gamma}\right)\frac{1}{\ln{\beta t}},
\end{eqnarray}
where $S_{\bf 0}(\infty)=1$.

For the other individuals we proceed similarly and from (\ref{Wat}), with $r\gg 1$, we have:
\begin{eqnarray}
\label{eq.21}  
{\hat I}_{ \bf r}(s, \beta)
\xrightarrow[s\rightarrow 0]{}
\frac{1}{2\pi\beta}K_{0}\left(r\sqrt{\frac{s}{\beta}}\right),
\end{eqnarray}
where  $K_{0}(x)$ is the usual modified Bessel
 function of the third kind \cite{Abramowitz}.
Using the small argument expansion of such a
 Bessel function \cite{Abramowitz}, 
we find that the long-time behavior in the region where  $\beta t\gg r^{2}\gg 1$ is

\begin{eqnarray}
\label{eq.23} 
S_{\bf r}(t) -
S_{\bf r}(\infty) \sim -
\frac{\ln{r^{2}}}{\ln{\beta t}}, 
\end{eqnarray}
where the stationary magnetization corresponds again to the unanimous opinion, as in $1d$: 
$ S_{\bf r}(\infty) = 1$.

The regime where $r^{2}\propto \beta t\rightarrow \infty$ should 
still be discussed separately.
In fact, in this regime, from  (\ref{eq.11}), (\ref{eq.20}) and (\ref{eq.21}) we 
have : 
\begin{eqnarray}
\label{eq.23.0} 
\hat{S}_{\bf r}(s) \xrightarrow[s\rightarrow 0,\, r\rightarrow \infty ]{}
-\frac{2K_0\left(r\sqrt{
s/\beta}
\right)}{s \ln{
\left(s/\beta
\right)}},
\end{eqnarray}
where $s\rightarrow 0$ and $r\rightarrow \infty$, but $r\sqrt{s/\beta}$ is finite.

\vspace{0.2cm}

From results (\ref{eq.16}), (\ref{eq.18}) and (\ref{eq.23}), (\ref{eq.23.0})
we infer that in low-dimensions, at large time, the effect of the zealot  appears at two length scales: 
(i) The opinion of individuals ``close'' ($r^2 \ll \beta t $) to the zealot evolves
 algebraically ($\propto t^{-1/2}$) in $1d$ and
logarithmically ($\propto 1/\ln{t}$) in $2d$ towards the unanimous opinion $S_{\bf r}(\infty)=1$.
 (ii) For individuals ``far'' ($r^2 \propto \beta t\rightarrow \infty $) from
 the zealot, the local magnetization evolves as a smooth scaling  function of
 $u\equiv r^{2}/2\sqrt{\beta t}$ 
in $1d$. This is however no longer the case in $2d$, where, due to
logarithmic terms arising 
in (\ref{eq.23.0}), the
 magnetization has not a  scaling form.
 A qualitatively  similar result, but within a different context, has recently been
 reported in Reference \cite{Majumdar}. One can
 also notice that results (\ref{eq.16}), (\ref{eq.18}) and (\ref{eq.23}),
 (\ref{eq.23.0}) are (mainly) independent of the rate $\gamma$: for 
the long-time behavior of the local magnetization; in low-dimensions, only
the fact that there is a biased individual (i.e. $\gamma>0$) matters.

 We therefore conclude that both in $1d$ and $2d$ the
 zealot eventually affects all the individuals: the number of voters having a
 final $+1$ opinion are within a ``circle'' whose radius increases as 
$\sqrt{\beta t}\rightarrow \infty$.

We now study the three-dimensional situation, and then consider its generalization
 to the case where $d\geq 3$.
 
When  $d\geq 3$, the integrals (\ref{Wat}) are well defined for all the values of $s$, and in particular when $s\rightarrow 0$. Therefore, conversely to what happens in $1d$ and $2d$, to determine the long-time behavior of the
 magnetization we cannot simply focus on the $q\rightarrow 0 $ expansion of (\ref{Wat}).
Fortunately, very recently Glasser and Boersma have been able to explicitly compute (\ref{Wat}) in the $3d$ case where $s=0$ \cite{Glasser}. We now take advantage of these findings to compute  the  stationary magnetization in $3d$. We therefore introduce a triplet
 $(a_{{\bf r}}, b_{{\bf r}}, c_{{\bf r}})$ of rational numbers 
depending on ${\bf r}$, given in Table 2 of Reference \cite{Glasser}, and the quantity $g_0\equiv\left(\frac{\sqrt{3}-1}{96\pi^{3}}\right)\Gamma^{2}
\left(\frac{1}{24}\right)\Gamma^{2}\left(\frac{11}{24}\right)$, where $\Gamma(z)$ is Euler's Gamma function \cite{Abramowitz}.  
  
With help of  the results obtained in \cite{Glasser}, we are in position to compute the
$3d$ stationary local magnetization (SLM) by taking the $s\rightarrow 0$ limit in (\ref{eq.11}):
\begin{eqnarray}
\label{eq.32}  
S_{\bf r}(\infty)  =
\frac{\gamma\left[
a_{\bf r} g_0^2 + c_{\bf r} g_0 +\frac{b_{\bf r} }{\pi^{2}}
\right]}{g_0\left[2\beta+\gamma g_0 \right]},
\end{eqnarray}
In particular, for ${\bf r}={\bf 0}$ we have the triplet 
$(a_{\bf 0}, b_{\bf 0},c_{\bf 0})=(1, 0, 0)$ \cite{Glasser} and thus obtain $S_{\bf 0}(\infty)= \frac{\gamma g_0}{2\beta +\gamma g_0}$.

To gain, in a simple manner, some further insight of the behavior of the discrete 
 expression (\ref{eq.32}) 
when $r\rightarrow \infty$, it turns out to be fruitful to take the {\it continuum 
limit} of (\ref{MS}). In this limit the SLM is ${\cal S}({\bf r},\infty)$, 
and we have to solve the problem of determining the electric potential due to a ``charge'' at the origin. 
One should pay due attention to the fact that this electrostatic
reformulation needs to be supplemented by additional information since, from the continuum limit 
 of (\ref{MS}), the charge is {\it a priori} an unknown quantity.
 To overcome this difficulty, in $3d$, we take advantage 
of our knowledge of the discrete version of the problem and,  with (\ref{eq.32}), we 
compute the charge at site ${\bf r}={\bf 0}$
 as $\frac{\gamma}{\beta}
[1-S_{\bf 0}(\infty)]=\frac{2\gamma}{2\beta +\gamma g_0}$.
 We therefore obtain the following $3d$ continuum stationary  equation:
\begin{eqnarray}
\label{CL}  
\Delta  {\cal S}({\bf r},\infty) =
-\frac{2\gamma}{2\beta+\gamma g_0}\,\delta^{3}({\bf r}),
\end{eqnarray}
where $\Delta$ is the $3d$-Laplacian operator and $\delta^{3}({\bf r})$ denotes 
the $3d$-Dirac delta function.

 The solution of Eq. (\ref{CL}) depends on $\gamma$ and reads:
\begin{eqnarray}
\label{CLsol}  
 {\cal S}({\bf r},\infty) =
\frac{\gamma}{2\pi (2\beta +\gamma g_0)} \, \frac{1}{r}\; \hspace{1cm} (r>0).
\end{eqnarray}
Comparing the predictions of the results (\ref{eq.32}) and (\ref{CLsol}),
 we notice that they agree very well, even for finite
 $r$: For instance, when $\gamma=2\beta$, at ${\bf r}=(3,1,0)$, we
have $ S_{\bf r}(\infty)=0.0339$ and ${\cal S}({\bf r},\infty)=0.0334$, whereas for ${\bf r}=(3,1,1)$
 $ S_{\bf r}(\infty)=0.0319$ and ${\cal S}({\bf r},\infty)=0.0319$.

Results (\ref{eq.32}) and (\ref{CLsol}) show that in $3d$ 
(conversely to what happens in low-dimensions) the SLM  is an isotropic
 (which is clear from (\ref{Wat}) and (\ref{CLsol})) but {\it non-uniform
  function} decaying  with the  norm of $ {\bf r}$: $S_{ \bf r}
(\infty) \approx {\cal S}({\bf r},\infty)= \frac{A_{3}(\gamma)}{r}$, where the
amplitude is  given by (\ref{CLsol}) [and $r>0$].
The reasoning can be extended to dimensions $d> 3$, where the electrostatic 
reformulation gives the result (for $r>0$):  
$ {\cal S}({\bf r},\infty)= \frac{A_{d}}{r^{d-2}}$. Again the computation of the amplitude 
$A_{d}$ requires the explicit knowledge of $\hat{I}_{\bf 0}(s=0,\beta)$ in dimensions $d>3$.

Despite the fact that in $d\geq 3$ the $s\rightarrow 0$ analysis of (\ref{Wat}) is a 
difficult task, we can infer (for $r$ finite) the long-time behavior from (\ref{eq.7}):
 $ S_{\bf r}(t) - 
S_{\bf r}(\infty)  \sim -(\beta t)^{1-\frac{d}{2}}$.

From Eq.(\ref{eq.7}) we can also compute the total
 magnetization: $M(t)\equiv \sum_{\bf r}S_{\bf r}(t)= M(0)+ 
\gamma\int_{0}^{t}d\tau \left[1-S_{\bf 0}(\tau)\right] $.
This expression shows that in the voter model ($\gamma=0$) the
 quantity $M(t)$ is conserved,
 which is no longer the case when $\gamma> 0$. With the help 
of the results (\ref{eq.15.0}), (\ref{eq.22}) and (\ref{eq.32}), we obtain the
 long-time behavior of $M(t)$ [here $M(0)=0$]:
\begin{center}
\begin{eqnarray}
\label{TotM123d}
M(t)\sim
\left\{
\begin{array}{l l l }
 (\beta t)^{1/2}
&\mbox{, \; $d=1$  } \\
\beta t/\ln{t}
&\mbox{, \; $d=2$} \hspace{1.0cm} (\beta t \rightarrow \infty)\\
\beta  t
&\mbox{, \;  $d\geq3$}.
\end{array}
\right.
\end{eqnarray}
\end{center}
These results show that in this inhomogeneous voter model
 the saturation time $t_s$, that is, the
 time necessary to have
$M(t_s)$ comparable to the size $L^{d}$  (where $L\rightarrow \infty$) of the system,
 scales as $\beta t_s \sim L^2$ in $1d$, $\beta t_s \sim L^{2}\ln{L}$ in $2d$,
 and $\beta t_s\sim L^{d}$ for $d\geq 3$. 
These statements are in agreement with results obtained for other models (see e.g.\cite{K1}). 

In this work we have developed a method to compute some exact properties of 
a class of stochastic many-body problems with inhomogeneities
and have explicitly presented this approach in the framework of an 
inhomogeneous voter  model
 where the usual voter dynamics is perturbed by the local presence of a
 single zealot.
 For this ``opinion dynamics'' problem, we have computed exactly, in
 dimensions $d=1, 2$ and $3$, the long-time magnetization (mean-average
 opinion of each voters). 
From our exact results we have seen that in 
low-dimensions the zealot (i.e. the inhomogeneity)
 always affects the mean-average opinion and that
 its effect propagates as $t^{1/2}$. In fact the mean opinion of individuals 
approaches, algebraically in $1d$ [see (\ref{eq.18})], according to the
 scaling expression (\ref{eq.16}),  and logarithmically slowly in $2d$ 
[see (\ref{eq.23})], the unanimity state favoured by the zealot.
These results are (mainly) independent of the strength of the biased 
individual.
 In $3d$ the situation is completely different and the (stable) stationary 
mean-average opinion of a voter is no longer uniform but follows the
 non-trivial isotropic function (\ref{eq.32}) that decays with 
the inverse of the distance to
 the zealot [see (\ref{CLsol})], a result also
 obtained in the continuum limit via a suitable electrostatic
 reformulation, and then extended to the case where $d>3$.
The findings  obtained in this work, and the  differences between 
the  behaviors observed in low-dimensions and in $d\geq 3$, can be qualitatively understood in realizing that the local magnetization is the solution of a diffusion-like equation (\ref{MS}) supplemented by a local boundary term (the zealot) and taking into account the fact that in $1d$ and $2d$ random-walks are recurrent (which, in our case, implies that all individuals interact with the zealot), while in $d\geq 3$ they are transient, and therefore there is a finite probability that individuals never interact with the zealot \cite{Privman,Liggett}. 
It is instructive to compare our results to those obtained in the
conventional voter model \cite{K1,Frachebourg,Ben,Dornic}:
 the presence of the zealot clearly
implies that the magnetization is no longer conserved and that 
the dynamics is not translationally invariant. This comparison 
also shows that a single inhomogeneity (here, the zealot) can deeply affect 
the stationary and the long-time properties of an interacting spin
 system, whose perturbed dynamical behavior depends on the dimension $d$.

Despite the fact that our method and results have been formulated 
in an ``opinion dynamics'' language, we 
 emphasize that they have a broad physical relevance and  can be applied
to a large class of stochastic many-body problems.
As a physical illustration we can consider the kinetics of the
 monomer-monomer catalysis on an inhomogeneous substrate
that locally desorps preferentially one species of 
monomer \cite{K1,Frachebourg,prep}.
Identifying spins $+1$ ($-1$)  with $A$ ($B$) particles, the
 monomer-monomer reaction
 can be mapped onto an Ising model with mixed voter 
and Kawasaki dynamics  \cite{K1,Frachebourg}, whose time-scales are
respectively defined 
by $1/\beta d$ and $1/\beta' d$, and an inhomogeneous
 term [as in (\ref{SF})] that mimics the {\it local}  
desorption  (at ${\bf r}={\bf 0}$), with rate $\gamma$, of
 particles of species $B$  \cite{K1,prep}. 
For this model, with ${\widetilde \beta}\equiv\beta + \beta'$, the local
 concentration $c_{{\bf r}}(t)$ of  $A$ particles obeys  to 
$(d/({\widetilde \beta} dt)) c_{\bf r}(t) = \Delta_{{\bf r}}c_{\bf r}(t)
-(\gamma/{\widetilde \beta})  (c_{\bf 0}(t)-1)\delta_{{\bf r, 0}}$, whereas 
the concentration of $B$ particles, at site ${\bf r}$, is 
 $1-c_{{\bf r}}(t) $ \cite{prep}. 
Comparing the equation for $c_{{\bf r}}$ with
Eq. (\ref{MS}), it is clear that the concentration of particles in the
 catalysis problem can immediately be inferred from the above
results for $S_{{\bf r}}(t)$ \cite{prep}.
Another field of pertinence of this work,
 according to the  well-known relation between the ($1d$) voter and 
 Glauber-Ising models \cite{Glauber,K1,Frachebourg}, is the area of inhomogeneous
 magnetism. 
We can also mention that our method, which can take into
account  the presence of many inhomogeneities and 
can deal with systems of more than two states per site, is
 of direct relevance for a large  class of reaction-diffusion 
models \cite{Privman}. In this context, a diffusive model of 
interacting particles where
 an homogeneous source is in competition with a local trap has 
been solved \cite{prep}.

We thank P.L. Krapivsky, S. Redner and  L. Zuppiroli for
 valuable discussions and reading the manuscript. Financial
 support of the vice-president (VPF) of the EPFL 
 and of the Swiss National Science Foundation under
 the fellowship 81EL-68473 are acknowledged.


\begin{thebibliography}{99}
\bibitem{Privman}
{\it Nonequilibrium Statistical Mechanics in One Dimension}, edited by 
V. Privman (Cambridge University Press, Cambridge, 1997); 
D. ben-Avraham and S. Havlin, {\it Diffusion and Reactions
 in Fractals and Disordered Systems} (Cambridge University Press, Cambridge, 2000);
S. Redner, {\it A guide to first-passage processes} (Cambridge University
Press, Cambridge, 2001); M. Henkel, E. Orlandini, and J. Santos,
Ann. Phys. (N.Y), {\bf 259}, 163 (1997); 
D. C. Mattis and M. L. Glasser, Rev. Mod. Phys. {\bf 70}, 979 (1998).
%
%
\bibitem{Liggett}
T. M. Liggett,{\it Interacting Particles Systems}, New York, Springer  (1985).
%
%
\bibitem{dis}
A. J. Bray, Adv. Phys. {\bf 43}, 357 (1994);
J. Duran and R. Jullien, Phys. Rev. Lett. 
{\bf 80}, 3547 (1998)
S. N. Majumdar, D. S. Dean  and P. Grassberger,
 {\it ibid.} 
{\bf 86}, 2301 (2001);
E. Ben-Naim and P. L. Krapivsky, Eur. Phys. J. E {\bf 8}, 507 (2002).
%
\bibitem{K1}
P. L. Krapivsky, Phys. Rev. A {\bf 45}, 1067 (1992); 
P. L. Krapivsky, J. Phys. A {\bf 25}, 5831 (1992).
%
\bibitem{Frachebourg}
L. Frachebourg and P. L. Krapivsky, Phys. Rev. E {\bf 53}, R3009 (1996).
%
\bibitem{Ben}
E. Ben-Naim, L. Frachebourg and P. L. Krapivsky, Phys. Rev. E {\bf 53}, 3078 (1996).
%
\bibitem{Dornic}
I. Dornic, H. Chat\'e, J. Chav\'e and H. Hinrichsen, Phys. Rev. Lett. {\bf 87}, 045701 (2001).
%
\bibitem{Vazquez}
F. Vazquez, P. L. Krapivsky and S. Redner, J. Phys. A.{\bf 36}, L61 (2003).
%
\bibitem{Abramowitz}
M. Abramowitz and  I. Stegun, {\it Handbook of Mathematical Functions} (Dover, NY, 1965).
%
\bibitem{Majumdar}
R. Rajesh and S. N. Majumdar, Phys. Rev. E {\bf 62}, 3186 (2000).
%
%
\bibitem{Glasser}
M. L. Glasser and J. Boersma, J. Phys. A {\bf 33}, 5017 (2000). 
%
\bibitem{prep}
M. Mobilia, in preparation.
%
\bibitem{Glauber}
R. J. Glauber, J. Math. Phys. {\bf 4}, 294 (1963).
\end{thebibliography}
\end{document}